\documentclass[aps,showpacs,twocolumn]{revtex4}
\usepackage{epsfig}

\newcommand{\be}{\begin{equation}}
\newcommand{\ee}{\end{equation}}
\newcommand{\bea}{\begin{eqnarray}}
\newcommand{\beaa}{\begin{eqnarray*}}
\newcommand{\eea}{\end{eqnarray}}
\newcommand{\eeaa}{\end{eqnarray*}}

\begin{document}

\title{\bf\Large {Ferrimagnetism of $MnV_2O_4$ spinel}}

\author{Naoum Karchev\cite{byline}}

\affiliation{Department of Physics, University of Sofia, 1164 Sofia, Bulgaria}

\begin{abstract}
The spinel $MnV_2O_4$ is a two-sublattice ferrimagnet, with site $A$ occupied by the $Mn^{2+}$ ion and site B by the $V^{3+}$ ion.
The magnon of the system, the transversal fluctuation of the total magnetization, is a complicated mixture of the sublattice $A$ and $B$ transversal magnetic fluctuations. As a result, the magnons' fluctuations suppress in a different way the manganese and vanadium magnetic orders and one obtains two phases. At low temperature $(0,T^*)$ the magnetic orders of the $Mn$ and $V$ ions  contribute to the
magnetization of the system, while at the high temperature  $(T^*,T_N)$, the vanadium magnetic order is suppressed by magnon fluctuations, and only the manganese ions have non-zero spontaneous magnetization.
A modified spin-wave theory is developed to describe the two phases and to calculate the magnetization as a function of temperature. The anomalous $M(T)$ curve reproduces the experimentally obtained ZFC magnetization.

\end{abstract}

\maketitle

This Letter is inspired from the experimental measurements
of the ZFC magnetization of $MnV_2O_4$\cite{vanadium1,vanadium2,vanadium3}. The profile of the experimental curve reproduces the anomalous magnetization curve predicted by L. Neel \cite{Neel,Wolf} for ferrimagnets with equal sublattice spins. This stimulates to model the manganese vanadate  spinel in the spirit of the Neel's theory. By comparing and contrasting ZFC and FC magnetization one gains insight into a magnetism of the manganese vanadate oxide.

The spinel $MnV_2O_4$ is a two-sublattice ferrimagnet, with site $A$ occupied by the $Mn^{2+}$ ion, which is in the $3d^5$ high-spin configuration with quenched orbital angular momentum, which can be regarded as a simple $s=5/2$ spin. The B site is occupied by the $V^{3+}$ ion, which takes the $3d^{2}$ high-spin configuration in the triply degenerate $t_{2g}$ orbital, and has orbital degrees of freedom. The measurements show that the set in of the magnetic order is at Neel temperature $T_N=56K$ \cite{vanadium1}, and that the magnetization has a maximum near $T^*=50K$. Below this temperature the magnetization sharply decreases and goes to zero when temperature approaches zero.

We consider a system which obtains its magnetic properties from $Mn$ and $V$ magnetic moments. It is shown that the true magnons
in this system, which are the transversal fluctuations corresponding to the total magnetization, are complicated mixtures of the $Mn$ and $V$ transversal fluctuations. The magnons interact with manganese and vanadium ions in a different way, and the magnons fluctuations
suppress the $Mn$ and $V$ ordered moments at different temperatures. As a result, the ferrimagnetic
phase is divided into two phases: low temperature phase $0<T<T^*$, where the magnetic orders of the $Mn$ and $V$ ions  contribute to the
magnetization of the system, and high temperature  phase $(T^*,T_N)$, where the vanadium magnetic order is suppressed by magnon fluctuations, and only the manganese ions have non-zero spontaneous magnetization. A modified spin-wave theory is developed to describe the two phases and to calculate the magnetization as a function of temperature. The anomalous $M(T)$ curve reproduces the experimentally obtained ZFC magnetization \cite{vanadium2,vanadium3}.

 Because of the strong spin-orbital interaction it is convenient to consider $jj$ coupling with $\textbf{J}^A=\textbf{S}^A$ and $\textbf{J}^B=\textbf{L}^B+\textbf{S}^B$. The sublattice $A$ total angular momentum is $j_A=s_A=5/2$, while the sublattice $B$ total angular momentum is $j_B=l_B+s_B$, and sublattice $A$ magnetic order is antiparallel with sublattice $B$ one.  In the simplest case one can consider $l_B=1$ and $s_B=3/2$. Then, the g-factor for the sublattice $A$ is $g_A=2$, and the atomic value of the $g_B$  is $g_B=1.6$. The saturated magnetization is $\sigma=2 \frac 52-1.6 \frac 52$. The experimental curve shows that the magnetization goes to zero when the temperature approaches zero, which indicates that the real value of $g_B$ is not the atomic
one but $g_B\approx 2$. The deviation is due to the anisotropy which increases the $g_B$-factor \cite{Wolf}.
The Hamiltonian of the system is
\bea \label{MnV1}
 H & = & - \kappa_A\sum\limits_{\ll ij \gg _A } {{\bf J}^A_{i}
\cdot {\bf J}^A_{j}}\,-\,\kappa_B\sum\limits_{\ll ij \gg _B } {{\bf J}^B_{i}
\cdot {\bf J}^B_{j}}\nonumber \\
& & +\,\kappa \sum\limits_{\langle ij \rangle} {{\bf J}^A_{i}
\cdot {\bf J}^B_{j}}\eea where the sums are over all sites of a three-dimensional cubic lattice:
$\langle i,j\rangle$ denotes the sum over the nearest neighbors, $\ll i,j \gg _{A(B)}$ denotes the sum over the sites of A(B) sublattice.
The first two terms  describe the ferromagnetic Heisenberg intra-sublattice
exchange $\kappa_A>0, \kappa_B>0$, while the third term describes the inter-sublattice exchange which is antiferromagnetic $\kappa>0$.

To proceed we use the Holstein-Primakoff representation of the total angular momentum vectors ${\bf J}^A_{j}(a^+_j,a_j)$ and ${\bf J}^B_{j}(b^+_j,\,b_j)$, where $a^+_j,\,a_j$
and $b^+_j,\,b_j$ are Bose fields. In terms of these fields and keeping only the quadratic and quartic terms, the effective Hamiltonian
Eq.(\ref{MnV1}) adopts the form , $H=H_2+H_4$ where
\bea\label{MnV2}
 H_2 & = & j_A \kappa_A\sum\limits_{\ll ij \gg _A }\left( a^+_i a_i\,+\,a^+_j a_j\,-\,a^+_j a_i\,-\,a^+_i a_j\right) \nonumber \\
  & + & j_B \kappa_B\sum\limits_{\ll ij \gg _B }\left( b^+_i b_i\,+\,b^+_j b_j\,-\,b^+_j b_i\,-\,b^+_i b_j\right) \\
 & + & \kappa \sum\limits_{\langle ij \rangle}\left[j_A b^+_j b_j + j_B a^+_i a_i -
 \sqrt{j_Aj_B}\left( a^+_i b^+_j+a_i b_j \right)\right] \nonumber
 \eea
 \bea\label{MnV3}
 H_4 & = & \frac 14 \kappa_A\sum\limits_{\ll ij \gg _A }\left[a^+_i a^+_j( a_i-a_j)^2 + (a^+_i- a^+_j)^2  a_i a_j\right] \nonumber \\
 & + & \frac 14 \kappa_B\sum\limits_{\ll ij \gg _B }\left[b^+_i b^+_j( b_i-b_j)^2 + (b^+_i- b^+_j)^2  b_i b_j\right]\nonumber \\
 & + & \frac 14 \kappa \sum\limits_{\langle ij \rangle}\left[
 \sqrt{\frac {j_A}{j_B}}\left( a_i b^+_j b_j b_j+a^+_i b^+_j b^+_j b_j \right)\right. \\
& + &\left. \sqrt{\frac {j_B}{j_A}}\left(a^+_i a_i a_i b_j+a^+_i a^+_i a_i b^+_j \right) - 4 a^+_i a_i b^+_j b_j \right] \nonumber
 \eea
and the terms without operators are dropped.

The next step is to represent the Hamiltonian in Hartree-Fock  approximation $H\approx H_{HF}=H_{cl}+H_q$ where
\bea\label{MnV4} H_{cl} & = & 12 N \kappa_A j_A^2 (u_A-1)^2+ 12 N \kappa_B j_B^2 (u_B-1)^2 \nonumber \\
& + & 6 N \kappa j_A j_B (u-1)^2
\eea
where $N=N_A=N_B$ is the number of sites on a sublattice.
 The Hamiltonian $H_q$ can be obtained from the Hamiltonian Eq.(\ref{MnV2}) replacing $\kappa_A$ with $\kappa_A u_A$, $\kappa_B$ with $\kappa_B u_B$ and $\kappa$ with $\kappa u$, where $u_A, u_B$ and $u$ are Hartree-Fock parameters, to be determined self-consistently. It is convenient to rewrite the Hamiltonian in momentum space representation
\be\label{MnV5}
H_q = \sum\limits_{k\in B_r}\left [\varepsilon^a_k\,a_k^+a_k\,+\,\varepsilon^b_k\,b_k^+b_k\,-
 \,\gamma_k \left (a_k^+b_k^+ + b_k a_k \right )\,\right ],
\ee
where the wave vector $k$ runs over the reduced  first Brillouin zone $B_r$ of a
cubic lattice. The dispersions are given by equalities
\bea\label{MnV6}
\varepsilon^a_k & = & 4j_A\,\kappa_A\,u_A\varepsilon_k
\,+\,6j_B\,\kappa u \\
\varepsilon^b_k & = & 4j_B\,\kappa_B\,u_B \varepsilon_k \,+\,6j_A\,\kappa\,u \nonumber \\
\gamma_k & = & 2\kappa\,u\,\sqrt{j_A\,j_B}\,\left(\cos k_x +\cos k_y + \cos k_z \right)\nonumber\eea with
$\varepsilon_k = 6-\cos(k_x+k_y)-\cos(k_x-k_y) - \cos (k_x+k_z)
 - \cos(k_x-k_z)- \cos(k_y+k_z) - \cos(k_y-k_z)$.

To diagonalize the Hamiltonian one introduces new Bose fields $\alpha_k,\,\alpha_k^+,\,\beta_k,\,\beta_k^+$ by means of the
transformation
\be\label{MnV7}
a_k\,=u_k\,\alpha_k\,+\,v_k\,\beta^+_k,\quad
 b_k\,=\,u_k\,\beta_k\,+\,v_k\,\alpha^+_k \ee
with coefficients $u_k$ and $v_k$ real functions of the wave vector $k$
\be\label{MnV8}
u_k\,=\,\sqrt{\frac 12\,\left (\frac
{\varepsilon^a_k+\varepsilon^b_k}{\sqrt{(\varepsilon^a_k+\varepsilon^b_k)^2-4\gamma^2_k}}\,+\,1\right
)},\ee
$v_k\,=\,sign (\gamma_k)\,\sqrt{u_k^2-1}$.
The transformed Hamiltonian adopts the form \be
\label{MnV9} H_q = \sum\limits_{k\in B_r}\left
(E^{\alpha}_k\,\alpha_k^+\alpha_k\,+\,E^{\beta}_k\,\beta_k^+\beta_k\,+\,E^0_k\right),
\ee
with new dispersions $E^{\alpha}_k\,=\,E^{+}_k$,\quad $ E^{\beta}_k\,=\, E^{-}_k$ where
\be  \label{MnV10} E^{\pm}_k\,=\,\frac 12\,\left[\sqrt{(\varepsilon^a_k\,+\,\varepsilon^b_k)^2\,-\,4\gamma^2_k}\,\pm\,(\varepsilon^a_k\,-\,\varepsilon^b_k)\right]\ee
 and vacuum energy
\be\label{MnV11}
 E^{0}_k\,=\,\frac
12\,\left [
\sqrt{(\varepsilon^a_k\,+\,\varepsilon^b_k)^2\,-\,4\gamma^2_k}\,-\,\varepsilon^b_k\,-\,\varepsilon^a_k\right]\ee

To obtain the system of equations for the Hartree-Fock parameters we consider
the free energy of a system with Hamiltonian $H_{HF}$ Eqs.(\ref{MnV4},\ref{MnV9})
\bea\label{MnV12}
\mathcal{F} & = & 12\kappa_A j_A^2 (u_A-1)^2+ 12\kappa_B j_B^2 (u_B-1)^2  \nonumber \\
& + & 6\kappa j_A j_B (u-1)^2 + \frac 1N \sum\limits_{k\in B_r}E^{0}_k \\
& + & \frac 1N \sum\limits_{k\in B_r}\left[ \ln\left(1-e^{-\beta E^{\alpha}_k}\right)\,+\,\ln\left(1-e^{-\beta E^{\beta}_k}\right)\right].\nonumber\eea
Then the three equations $\partial\mathcal{F}/\partial u_A=0,\,\partial\mathcal{F}/\partial u_B=0,\,$ and $\partial\mathcal{F}/\partial u=0$ adopt the form
\bea\label{MnV13} u_1 & = & 1-\frac {1}{6j_1} \frac 1N \sum\limits_{k\in B_r} \varepsilon_k \left[u_k^2 \,n_k^{\alpha}\, +\, v_k^2\, n_k^{\beta}\, +\, v_k^2\right]\nonumber \\
u_2 & = & 1-\frac {1}{6j_2} \frac 1N \sum\limits_{k\in B_r} \varepsilon_k \left[v_k^2 \,n_k^{\alpha}\, +\, u_k^2\, n_k^{\beta}\, +\, v_k^2\right]\nonumber \\
u & = & 1-\frac 1N \sum\limits_{k\in B_r} \left[\frac {1}{2j_1}\left(u_k^2 \,n_k^{\alpha}\, +\, v_k^2\, n_k^{\beta}\, +\, v_k^2\right)\right. \\
& + & \left. \frac {1}{2j_2} \left(v_k^2 \,n_k^{\alpha}\, +\, u_k^2\, n_k^{\beta}\, +\, v_k^2\right)\right. \nonumber \\
& - & \left.\frac 23 \kappa u\left(1+n_k^{\alpha}+n_k^{\beta}\right) \frac {\left(\cos k_x +\cos k_y + \cos k_z \right)^2}{\sqrt{(\varepsilon^a_k\,+\,\varepsilon^b_k)^2\,-\,4\gamma^2_k}}
\right]\nonumber
\eea
where $n_k^{\alpha}$ and $n_k^{\beta}$ are the Bose functions of $\alpha$ and $\beta$ excitations. Hartree-Fock parameters, the solution of the system of equations (\ref{MnV13}), are positive function of $T/\kappa$, $u_1(T/\kappa)>0,\,u_2(T/\kappa)>0$ and $u(T/\kappa)>0$. Utilizing these functions, one can calculate the spontaneous magnetization $M^A=<J^A_3>$\, and\, $M^B=<J^B_3>$\, of $Mn$ and $V$ ions respectively. In terms of the Bose functions of the $\alpha$ and $\beta$ excitations they adopt the form
\bea\label{MnV14}
M^A & = & j_A\,-\,\frac 1N \sum\limits_{k\in B_r} \left[u_k^2 \,n_k^{\alpha}\, +\, v_k^2\, n_k^{\beta}\, +\, v_k^2\right] \\
M^B & = & - \,j_B\,+\,\frac 1N \sum\limits_{k\in B_r} \left[v_k^2 \,n_k^{\alpha}\, +\, u_k^2\, n_k^{\beta}\, +\, v_k^2\right]\nonumber \eea

The magnon excitations  in the effective theory are a complicated mixture of the sublattices' $A$ and $B$  transversal fluctuations. As a
result, the magnon fluctuations suppress in a different way $Mn$ and $V$ magnetic orders. Quantitatively, this depends on the coefficients
$u_k$ and $v_k$ in Eq.(\ref{MnV14}). At characteristic temperature $T^*$ $V$ spontaneous magnetization becomes equal to zero, while $Mn$ spontaneous magnetization is still nonzero. Above $T^*$ the system of equations (\ref{MnV13}) has no solution, and one has to modify the spin-wave theory.

Once suppressed, the magnetic moment of $V$ ions cannot be restored increasing the temperature above T*. To formulate this mathematically
we modify the spin-wave theory using the idea on description of paramagnetic phase of 2D ferromagnets ($T>0$) by means of modified spin-wave
theory \cite{Takahashi1,Takahashi2}. We consider two-sublattice system and to enforce the magnetic
moments on the two sublattices to be equal to zero in paramagnetic pase
we introduce two parameters $\lambda_1$ and $\lambda_2$ \cite{Karchev08a}. The new Hamiltonian is obtained from the old one Eq.(\ref{MnV1}) adding two new terms
\be
\label{MnV15} \hat{H}\,=\,H\,-\,\sum\limits_{i\in A}
\lambda_A J^A_{3i}\,+\,\sum\limits_{i\in B} \lambda_B J^B_{3i} \ee
In momentum space, the Hamiltonian adopts the form Eq.(\ref{MnV5}) with new dispersions $\hat{\varepsilon}^a_k=\varepsilon^a_k+\lambda_A$
and $\hat{\varepsilon}^b_k=\varepsilon^b_k+\lambda_B$, where the old dispersions are given by equalities (\ref{MnV6}). We utilize
the same transformation Eq.(\ref{MnV7}) with coefficients $\hat{u}_k$ and $\hat{v}_k$ which depend on the
new dispersions in the same way as the old ones depend on the old dispersions Eq.(\ref{MnV8}). In terms of the $\alpha_k$ and
$\beta_k$ bosons, the Hamiltonian $\hat{H}_q$  adopts the form Eq.(\ref{MnV9}) with dispersions $\hat{E}^{\alpha}_k$ and
$\hat{E}^{\beta}_k$, which can be written in the form Eq.(\ref{MnV10}) replacing $\varepsilon^a_k$ and $\varepsilon^b_k$ with
$\hat{\varepsilon}^a_k$ and $\hat{\varepsilon}^b_k$.

We have to do some assumptions for parameters $\lambda_A$ and $\lambda_B$ to ensure correct definition of the two-boson
theory. For that purpose, it is convenient to represent the parameters in the form
$\lambda_A\,=\,6 \kappa j_B (\mu_A\,-1) \quad\lambda_B\,=\,6 \kappa j_A (\mu_B\,-\,1)$.
In terms of the parameters $\mu_A$ and $\mu_B$, the dispersion reads
$\hat{\varepsilon}^a_k\,=\,4 j_A
\kappa_A\,\varepsilon_k\,+\,6\,\kappa\,j_B\mu_A\quad
\hat{\varepsilon}^b_k\,=\,4j_B\,\kappa_B\,\varepsilon_k\,+\,6\,\kappa\,j_A\mu_B$. The conventional spin-wave theory is reproduced when
$\mu_A=\mu_B=1$($\lambda_A=\lambda_b=0$). We assume $\mu_A$and $\mu_B$ to be positive ($\mu_A>0,\,\mu_B>0$). Then,
$\hat{\varepsilon}^a_k>0$, $\hat{\varepsilon}^b_k>0$, for all values of the wave-vector $k$ if the Hartree-Fock parameters are positive too.
The Bose theory is well defined if $E^{\alpha}_k\geq 0,\quad
E^{\beta}_k\geq 0$. This comes true if $\mu_A\mu_B\geq1$. In the case $\mu_A \mu_B>1$, both $\alpha_k$ and
$\beta_k$ bosons are gapped excitations. In the particular case, $\mu_A \mu_B=1$, long-range excitations (magnons) emerge in the
system.

We introduced the parameters $\lambda_A$ and $\lambda_B$ ($\mu_A, \mu_B$) to enforce the sublattice $A$ and $B$ spontaneous magnetizations
to be equal to zero in paramagnetic phase. We find out the parameters $\mu_A$ and $\mu_B$, as well as Hartree-Fock parameters, as functions of temperature, solving the system of five equations, equations (\ref{MnV13}) and the equations $M^A=M^B=0$, where the ordered moments have the same representation as Eq.(\ref{MnV14}) but with coefficients $\hat{u}_k,\,\,
\hat{v}_k$, and dispersions $\hat{E}^{\alpha}_k,\,\, \hat{E}^{\beta}_k$ in the expressions for the Bose functions.
The numerical calculations show that for high enough temperature
$\mu_A\mu_B>1$. When the temperature decreases the product $\mu_A\mu_B$ decreases remaining larger than one. The temperature at which the product becomes equal to one ($\mu_A\mu_B=1$) is the Neel temperature. Below $T_N$, the spectrum contains long-range (magnon) excitations, thereupon $\mu_A\mu_B=1$. It is convenient to represent the parameters in the following way:
\be\label{MnV16}\mu_A=\mu, \quad\quad \mu_B=1/\mu.\ee

In ordered phase magnon excitations are origin of suppression of the magnetization. Near the zero temperature their contribution is
small and at zero temperature $Mn$ and $V$ spontaneous magnetization reach their saturation. Increasing the temperature magnon fluctuations suppress the magnetic order of $Mn$ and $V$ ions in different way. At $T^*$ the $V$ spontaneous magnetization becomes equal to zero.
Increasing the temperature above $T^*$, the magnetic moment of the vanadium ions should be zero. This is why we impose the condition
$M^B(T)=0$ if $T>T^{*}$. For temperatures above $T^*$, the parameter $\mu$ and the Hartree-Fock parameters are solution of a system of four equations, equations (\ref{MnV13}) and the equation $M^B=0$. We utilize the
obtained function $\mu(T), u_A(T), u_B(T), u(T)$ to calculate the spontaneous magnetization $M^A$ of the
$Mn$ ions as a function of the temperature. Above $T^*$, $M^A(T)$ is equal to the magnetization of the system.

We consider two-sublattice ferrimagnet with $Mn$ ions on sublattice $A$ and $V$ ions on sublattice $B$. The sublattice $A$ total angular moment is $j_A=s_A=5/2$, and $g$-factor $g_A=2$. The sublattice $B$ total angular momentum is $j_B=l_B+s_B=5/2$, and $g$-factor $g_B=2$ which is larger then atomic value, because of the anisotropy. The magnetization of the system $g_A\,M^A\,+\,g_B\,M^B$ as a function of the temperature is depicted  in Fig.1 for parameters\, $\kappa_A/\kappa\,=\,0.6$\, and \,$\kappa_B/\kappa\,=\,0.01$
\begin{figure}[!ht]
\epsfxsize=8.5cm 
\epsfbox{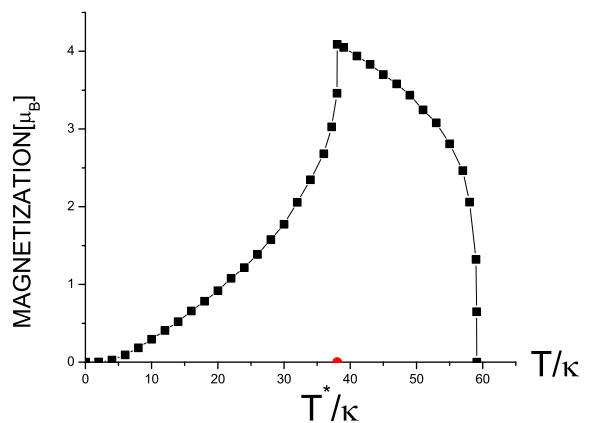} \caption{(color online)\,The magnetization as a function of $T/\kappa$ for parameters\, $\kappa_A/\kappa\,=\,0.6$\, and \,$\kappa_B/\kappa\,=\,0.01$}\label{fig1}
\end{figure}

The figure is in a very good agrement with the experimental ZFC magnetization curves \cite{vanadium2,vanadium3}.
(see Fig.1 \cite{2fmp6}).
The anomalous temperature dependence of the magnetization, predicted by Neel, is reproduced, but there is an
important difference between Neel's theory, interpretation of NMR results  \cite{vanadium2,vanadium3}, and the present modified spin-wave theory. Neel's calculations predict a temperature $T_N$ at which both the sublattice $A$ and $B$ magnetization become equal to zero. The modified spin wave theory predicts to phases: at low temperatures $(0,T^*)$ $Mn$ and $V$ magnetic orders contribute to the
magnetization of the system, while at  high temperatures  $(T^*,T_N)$ only $Mn$ ions have non-zero spontaneous magnetization.
The vanadium electrons start to form magnetic moment at $T^*$, and an evidence for this is the abrupt decrease of magnetization below $T^*$, which also indicates that the magnetic order of vanadium electrons is anti-parallel with the order of $Mn$ electrons.

Two ferromagnetic phases where theoretically predicted, very recently,  in spin-Fermion systems, which obtain their magnetic properties
from a system of localized magnetic moments being coupled to conducting electrons \cite{Karchev08a}. At the characteristic
temperature $T^{*}$, the magnetization of itinerant electrons becomes zero, and high temperature ferromagnetic phase ($T^{*}<T<T_C$) is a
phase where only localized electrons give contribution to the magnetization of the system. An anomalous increasing of
magnetization below $T^{*}$ is obtained in good agrement with experimental measurements of the ferromagnetic phase of $UGe_2$\cite{2fmp6}.

The results of the present paper and the previous one \cite{Karchev08a} suggest that $T^*$ transition from  a magnetic phase to another magnetic phase is a generic feature of the two magnetic orders systems. The additional phase transition demonstrates itself through the anomalous temperature variation of the spontaneous magnetization, but it is important to discuss alternative experimental detections of $T^*$ transition. This is why we consider the FC magnetization curves \cite{vanadium2,vanadium3}.
For samples cooled in a field (FC magnetization) the field leads to formation of a single domain and, in addition, increases the chaotic order of the spontaneous magnetization of the vanadium electrons, which is antiparallel with it. As a result the average value of the vanadium magnetic order decreases and does not compensate the $Mn$ magnetic order. The magnetization curves depend on the applied field, and does not go to zero. For a larger field the (FC) curve increases when temperature decreases below Neel temperature . It has a maximum at the same temperature $T^*<T_N$ as the ZFC magnetization, and a minimum at $T_1^*<T^*$. Below $T_1^*$  the magnetization increases monotonically when temperature approaches zero.

The experiments with samples cooled in field (FC magnetization) provide a new opportunity to clarify the magnetism of the manganese vanadium oxide spinel. The applied field is antiparallel with vanadium magnetic moment and strongly effect it. On the other hand the experiments show that there is no difference between ZFC and FC magnetization curves when the temperature runs the interval ($T^*,T_N$) \cite{vanadium2,vanadium3}. They begin to diverge when the temperature is below $T^*$. This is in accordance with the theoretical prediction that the vanadium magnetic moment does not contribute the magnetization when $T>T^*$, and $T^*$ is the temperature at which the vanadium ions start to form magnetic order.
Because of the strong field, the vanadium bands are split and part of the magnetic orders are reoriented to be parallel with field and magnetic order of $Mn$ electrons. The description of this case is more complicate and requires three magnetic orders to be involved. When $T^*<T<T_N$ only $Mn$ ions have non zero spontaneous magnetization. At $T^*$ vanadium magnetic order antiparallel with magnetic order of $Mn$ sets in and partially compensates it. Below $T_1^*$ the reoriented magnetic orders give contribution, which explains the increasing of the magnetization of the system when the temperature approaches zero.

To conclude, we note that a series of experiments with different applied field could be decisive for the confirmation or rejection of the $T^*$ transition. Increasing the applied field one expects increasing of $T^*_1$ and when the field is strong enough, so that all vanadium electrons are reoriented, an anomalous increasing of magnetization below $T^{*}$ would be obtained as within the ferromagnetic
phase of $UGe_2$ \cite{2fmp6}.

\end{document}